\documentclass[aps,prl,twocolumn,superscriptaddress]{revtex4-1}

\usepackage{fancyhdr} 

\usepackage{hyperref} 

\usepackage{pdfpages} 
\makeatletter
\AtBeginDocument{\let\LS@rot\@undefined}
\makeatother

\usepackage{verbatim} 
\usepackage{amsmath} 
\usepackage{amsthm} 
\usepackage{amssymb}	
\usepackage{graphicx} 
\usepackage[dvips,letterpaper,margin=0.75in,bottom=0.5in]{geometry}

\let\baraccent=\= 
\renewcommand{\=}[1]{\stackrel{#1}{=}} 

\newcommand{\gae}{\lower 2pt \hbox{$\,
    \buildrel{\scriptstyle >}\over {\scriptstyle \sim}\,$}}
\newcommand{\lae}{\lower 2pt \hbox{$\,
    \buildrel{\scriptstyle <}\over {\scriptstyle \sim}\,$}}

\newcommand{\sinc}{\mathrm{sinc}}

\newcommand{\abs}[1]{\left| #1 \right|} 
 
\usepackage{color}
\usepackage{ulem}

\begin{document}

\title{
Energy drag in particle-hole symmetric systems as a quantum quench
}

\author{William Berdanier}
\email[]{wberdanier@berkeley.edu}
\affiliation{Department of Physics, University of California, Berkeley, CA 94720, USA}

\author{Thomas Scaffidi}
\affiliation{Department of Physics, University of California, Berkeley, CA 94720, USA}
\affiliation{Department of Physics, University of Toronto, Toronto, Ontario, M5S 1A7, Canada}

\author{Joel E. Moore}
\affiliation{Department of Physics, University of California, Berkeley, CA 94720, USA}
\affiliation{Materials Sciences Division, Lawrence Berkeley National Laboratory, Berkeley, CA 94720, USA}

\date{\today}

\begin{abstract}
Two conducting quantum systems coupled only via interactions can exhibit the phenomenon of Coulomb drag, in which a current passed through one layer can pull a current along in the other. However, in systems with particle-hole symmetry -- for instance, the half-filled Hubbard model or graphene near the Dirac point -- the Coulomb drag effect vanishes to leading order in the interaction. Its {\it thermal} analogue, whereby a thermal current in one layer pulls a thermal current in the other, does not vanish and is indeed the dominant form of drag in particle-hole symmetric systems. By studying a quantum quench, we show that thermal drag, unlike charge drag, displays a non-Fermi's Golden Rule growth at short times due to a logarithmic scattering singularity generic to one dimension. Exploiting the integrability of the Hubbard model, we obtain the long-time limit of the quench for weak interactions. Finally, we comment on thermal drag effects in higher dimensional systems.
\end{abstract}

\maketitle

Since its inception~\cite{pogrebinskii}, the Coulomb drag phenomenon -- whereby a charge current in one layer pulls a reciprocal current in another through Coulomb interactions alone -- has shed light on the special role of interaction effects in quantum transport~\cite{RevModPhys.88.025003}. Coulomb drag measurements have been instrumental in studying the microscopic structure of systems as diverse as double-quantum well structures~\cite{PhysRevLett.66.1216,EISENSTEIN1992107}, excitons in electron-hole bilayers~\cite{doi:10.1063/1.2132071,PhysRevLett.101.246801,PhysRevLett.102.026804,PhysRevLett.106.236807}, quantum Hall states~\cite{PhysRevB.49.11484,PhysRevLett.88.126804,PhysRevLett.93.036801,PhysRevLett.84.5808,doi:10.1116/1.3319260}, Luttinger liquids~\cite{Debray_2001,Laroche631}, spin currents in two-dimensional electron gases~\cite{PhysRevB.62.4853,Weber:2005aa}, and bilayer graphene~\cite{KIM20121283,PhysRevB.83.161401,Gorbachev:2012aa,PhysRevB.76.081401,PhysRevLett.109.236602,PhysRevLett.110.026601,PhysRevLett.111.166601}, among others. From the theoretical point of view, the Coulomb drag conductivity generally shows a rich dependence with temperature, with each regime dominated by different microscopic processes, and has been generalized in many directions~\cite{RevModPhys.88.025003}. Given the recent interest in the hydrodynamic behavior of electrons in solids~\cite{PhysRevLett.94.111601,PhysRevLett.118.226601,Mackenzie_2017}, an analogy can also be made between the Coulomb drag and the shear viscosity, two processes leading to the equalization of currents in neighboring layers.


In light of this history, it stands to reason that the Coulomb drag effect between {\it thermal} currents, first studied to our knowledge in Ref.~\cite{PhysRevB.98.035415} -- in which a thermal current in one layer may drag along a reciprocal thermal current in the other through Coulomb interactions -- could elucidate the microscopic structure of quantum systems as well. In fact, in one particularly interesting class of quantum systems -- those having particle-hole symmetry -- Coulomb charge drag effects are known to vanish at leading order~\cite{PhysRevB.76.081401}.  Momentum is transferred between the layers at this order, but it cannot result in a charge current~\cite{Rojo_1999}.  This is not a straightforward effect of symmetry, which would lead to vanishing at all orders; rather the leading process in perturbation theory is independent of the sign of the scattering potential, as with the Born approximation, so that the currents induced by particle-particle and particle-hole scattering cancel.  Such systems are prime candidates for the study of thermal drag, as thermal drag need not vanish under particle-hole symmetry, and we find it to be the dominant form of drag in such systems. Examples include the Hubbard model at half filling, graphene near the Dirac point, and superconductors probed at low energy, among others.


\begin{figure}
    \centering
    \includegraphics[width=0.9\columnwidth]{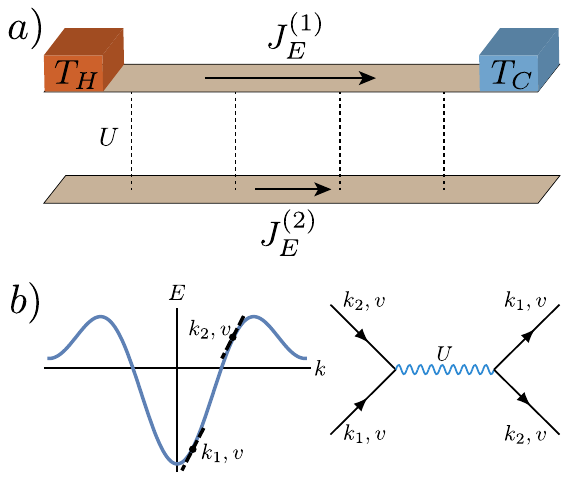}
    \caption{\label{fig:cartoon} (a) The thermal Coulomb drag geometry considered in this paper. A conducting quantum system's top layer is held at a temperature gradient by connecting it to two reservoirs at temperatures $T_H > T_C$, causing a thermal current to flow; through quantum interactions $U$, a thermal current is dragged in the bottom layer. (b) The source of the divergent scattering process leading to the breakdown of the usual Fermi's Golden Rule in one-dimensional systems, namely when all incoming and outgoing particles have the same velocity $v$ but differ in energy.}
\end{figure}

In this Letter, we focus on thermal drag between particle-hole symmetric quantum systems, viewed through the lens of a quantum quench of the inter-layer interactions in a bilayer system. We find that thermal drag does indeed dominate drag physics in these systems and, in sharp contrast to charge drag, suffers from a scattering singularity generic to one-dimensional band structures. This singularity leads to a violation of the na\"ive Fermi's Golden Rule, where the rate of change of the thermal current is {\it logarithmic} in time rather than constant, in the thermodynamic limit. This implies that a simple scattering rate analysis is generally incorrect, and more sophisticated perturbation theory analysis must be used; in particular, the approximation of linearizing the spectrum cannot be used when dealing with thermal currents without some method of regulation.

\paragraph{A quench and a Kubo formula.} To study the thermal drag, let us consider the paradigmatic one-dimensional Hubbard model,

\begin{equation}
    H = -t \sum_{\langle i j \rangle,\sigma} c_{i,\sigma}^\dagger c_{j,\sigma}+ U \sum_i (n_{i,\uparrow}-\frac12) (n_{i,\downarrow} - \frac12)
\end{equation}

where $\{c_{i,\sigma}^\dagger,c_{j,\sigma'}\} = \delta_{ij} \delta_{\sigma \sigma'}$. Let us view the two spin species as each forming separate quantum wires, with on-site interactions coupling them. {We note that the limit of on-site interactions can be physically motivated as originating from a screened Coulomb potential with small screening length.} Initialize one species, say spin-down, in a thermal state at temperature $T$ with some small initial energy current, and initialize the other spin species in a thermal state with no energy current (with $U=0$). Explicitly, since the free fermion chain may be diagonalized by a simple Fourier transform with energies $E_k = -2t \cos k$ and velocities $v_k = 2t \sin k$ (assuming periodic boundary conditions), such a state is given by 

\begin{equation}
\langle n_k^{\sigma} \rangle = \frac{1}{1+\exp(\beta(-2 t \cos k - \mu))} - \delta_{\sigma \downarrow}\epsilon \sin(2k),
\end{equation}

with $\epsilon$ a small parameter {ensuring the validity of linear response}. The charge and thermal current operators carried by the $\sigma$ spin species are given respectively by $J^\sigma = L^{-1} \sum_k v_k n_k^\sigma$ and $J_E^\sigma = L^{-1}\sum_k E_k v_k n_k^\sigma$, hence this initial state has $\langle J_E^\sigma \rangle = \epsilon\delta_{\sigma \downarrow}$ and $\langle J^\sigma \rangle=0$ (diagrammed in Fig.~\ref{fig:cartoon}(a)). In this setup, the spin-down channel is the ``drive'' layer and the spin-up channel is the ``response'' layer in the usual terminology of Coulomb drag, with the caveat that the ``drive'' current is allowed to relax (which does not change the short-time dynamics). We note that, while somewhat unorthodox, this quench interpretation of the Coulomb drag problem is physically reasonable and allows for the use of techniques from scattering theory and integrability that would be inapplicable in an equilibrium description. To avoid confusion, from now on, we set the Hubbard hopping paramter $t=1$.

At time $t=0$, let us quench on the interaction term $U$. We are interested in the change over time of the heat current in the spin-up channel. From the perspective of linear response, one would expect that an initial thermal current in the spin-down channel would drag along a thermal current in the spin-up channel, leading to the development of a temperature gradient for the spin-up species that is proportional to the initial energy current. This would give a thermal drag conductivity of 

\begin{equation}
\kappa_D = \frac{J_E^{(1)} }{\nabla T^{(2)} }
\end{equation}

where $J_E^{(1)}$ is taken at time $t=0$, and here $(1)$ refers to spin-up and $(2)$ to spin-down. Now, generally speaking, there is no perturbing Hamiltonian for a temperature gradient, so there is no straightforward method of deriving a Kubo formula for thermal conductivities. One may argue, however, based on entropy production in the system, that there exists an effective perturbing Hamiltonian and from this derive a Kubo formula~\cite{pottier}. Adapting this method, we arrive at a Kubo formula for the thermal drag conductivity~\cite{PhysRevB.98.035415}\footnote{See the Supplemental Material, section I for a derivation of the thermal drag Kubo formula, which includes Refs.~\cite{doi:10.1143/JPSJ.12.1203,PhysRev.135.A1505,pottier,PhysRevB.78.205407,ziman_book,PhysRevB.78.205407,PhysRevB.73.165104}.},

\begin{equation}
\label{eq:Kubo}
\kappa_{ab}^{\sigma\sigma'}(q,\omega) = \frac{1}{VT} \int_0^\infty dt e^{(i\omega - 0^+)t} \int_0^\beta d\lambda \langle J_{Q,b}^\sigma(-q,-i\lambda) J_{Q,a}^{\sigma'}(q,t)\rangle
\end{equation}
with $V$ the system size, $\sigma$ and $\sigma'$ layer indices, $q$ the wavevector, $J_Q$ the heat current, and $a$ and $b$ spatial indices (in the case of higher dimensional systems). 

With this Kubo formula in hand, we can connect our quench picture to the thermal drag conductivity by the following argument: if the initial rate of change of the energy current in the spin-down species is some rate $\partial_t \langle J_E^\uparrow \rangle = \Gamma$, then by the fluctuation-dissipation theorem~\cite{Kubo_1966} we should expect that the two-point function is exponentially decaying with the same rate $\Gamma$. This would give $\kappa_D \sim \int_0^\infty dt e^{i\omega t} e^{-\Gamma t} = 1/(\Gamma - i \omega)$, which, identifying $\Gamma = 1/\tau$ with $\tau$ a scattering time, would reproduce the usual Drude relation. We caution that in this case, however, a na\"ive Drude analysis will fail due to the complicated behavior of the energy current post-quench, which we examine below.

To calculate $\Gamma$, we seek the quantity $\partial_t n_k^\uparrow$, under the perturbation of the Hubbard interaction. To lowest (second) order in $U$,

\begin{equation}
\label{eq:dtnk}
\partial_t n_k^\uparrow = U^2 \sum_{k_2,k_3,k_4}S_{k k_2}^{k_3 k_4} \frac{\sin(t \Delta E)}{\Delta E} \delta(\Delta k),
\end{equation}

where $S_{k k_2}^{k_3 k_4} = (1-n_k^\uparrow) (1-n_{k_2}^\downarrow)n_{k_3}^\uparrow n_{k_4}^\downarrow - n_k^\uparrow n_{k_2}^\downarrow (1-n_{k_3}^\uparrow)(1-n_{k_4}^\downarrow)$ is the net Fermi factor for the inward and outward scattering processes, $\Delta k = k + k_2 - k_3 - k_4$ and $\Delta E = E_{k} + E_{k_2} - E_{k_3} - E_{k_4}$. In the usual Fermi's Golden Rule, one takes the limit of large $t$, which sends $\sin(t \Delta E)/\Delta E \to \pi \delta(\Delta E)$ provided that the quantity being integrated against does not diverge at $\Delta E = 0$. This is the case for Coulomb drag of charge currents, which is well-behaved; however, this is {not} the case for the energy current, as we shall see, and we must deal with the divergence carefully.

Imposing momentum conservation, the energy current grows as

\begin{equation}
\label{eq:dtJE}
\partial_t J_E^\uparrow = \frac{2}{L} \sum_k \sin(2k) \partial_t n_k^\uparrow.
\end{equation}

{We can usefully rewrite this expression by moving the sum on $k$ to an integral in energy space of a quantity $G(E)$, integrating against a kind of ``density of states''~\footnote{See the Supplemental Material, section II, for a full derivation of the logarithmic divergence encountered in perturbation theory.}. Focusing on half-filling $\mu=0$, the function $G(E)$ contains the essential divergence of the response energy current, namely

\begin{align}
G(E)& \propto  \int dk_1 dk_3 \sum_{\nu = 1,2}  \frac{F(k_1,k_{2,\nu},k_3)}{\abs{v(k_1 + k_{2,\nu} - k_3) - v(k_{2,\nu})}}
\end{align}

with $v(k) = \partial_k E(k) \propto \sin k$ the group velocity, the function $F$ does not diverge, and $\nu$ indexes the solutions to $\Delta E - E = 0$. Clearly, the source of the divergence is the difference of velocities in the denominator, corresponding to a resonance of points in $k$ space with different energies but the same velocity. Physically, this shows that the energy current operator diverges at small energies $\Delta E \approx 0$ which are directly probed by the $\sinc(t\Delta E )$ term in perturbation theory, and it is because of this singular behavior that Fermi's Golden Rule breaks down.}

There are two conditions under which the denominator diverges: the trivial case of $k_1=k_3$, and the nontrivial second solution. In the first instance, one can readily see that the numerator also vanishes, and hence there is no divergence. For the second solution, which occurs here at $k_1 + k_2 - k_3 = \pi - k_2$ but must occur somewhere in a generic one-dimensional band structure, one finds that the numerator also vanishes for a charge current -- and hence, it is well-behvaed -- while it does not for the energy current. The divergence is point-like, in the sense that for every incoming $k$ there is a finite set of partners $\{k'\}$ with the same velocity. That there must be at least one partner is a consequence of the lattice, i.e. the periodicity of the band structure (see Figure~\ref{fig:cartoon}(b)).

{At small but finite $E$, we can regularize the denominator, ultimately leading to a logarithmic divergence. A careful accounting yields }

\begin{equation}
g(E) = \epsilon \frac{4 U^2}{(2\pi)^3} \int_{-\pi}^\pi dk \frac{f(k)}{\abs{\sin k/2}} \log E
\end{equation}

where $g(E) =(G(E) + G(-E))/2$ is the symmetric part of $G(E)$, $f(k) = -2 \sin^2(k) n(E_k) n(-E_k)$, and $n(E)$ is the Fermi-Dirac distribution. Finally, using $\int_{-\infty}^\infty dx \ \log(x) \sinc(x t) = - \pi (\gamma + \log t) / t$, with $\gamma$ the Euler-Mascheroni constant, and keeping only the dominant term in the large $t$ limit, we arrive at the result

\begin{equation}
\frac{\partial_t J_E^\uparrow}{J_E^\downarrow(t=0)} = \alpha \log t + \mathcal{O}(1),
\end{equation}
with 
\begin{equation}
\label{eq:alpha}
\alpha(T) = \frac{U^2}{\pi^2} \int_{0}^{\pi} dk \abs{u(k)}^2 \frac{\sin^2 k \csc(k/2)}{1 + \cosh(2 \beta \sin(k/2))},
\end{equation}

 {where for generality, we have allowed for $k$-dependent interactions, $U(k) = U u(k)$, and in the Hubbard model with onsite interactions $u(k) = 1$. We remark that this logarithmic behavior is quite general: we expect it for any lattice band structure in 1D, as such band structures must generically have points where $v(k) = v(k')$ but $E(k) \not = E(k')$. Further, other kinds of interactions only modify the prefactor of the log growth.} This integral cannot be computed analytically, but for Hubbard, the low- and high-temperature limits are readily analyzed. First, at low temperatures, the denominator is a strongly peaked function about $k=0$; expanding the numerator in Taylor series and performing the integration yields

\begin{equation}
\alpha(T) \underset{T \ll 1} {\approx} \frac{4 U^2\log 2}{\pi^2} T^2,
\end{equation}

in units of Hubbard hopping $t=1$ and $k_B=1$. In the high temperature limit the demoninator is approximately constant, yielding

\begin{equation}
\alpha(T)  \underset{T \gg 1} {\approx} 4 U^2 / 3 \pi^2.
\end{equation}

\begin{figure}
	\centering
	\includegraphics[width=\columnwidth]{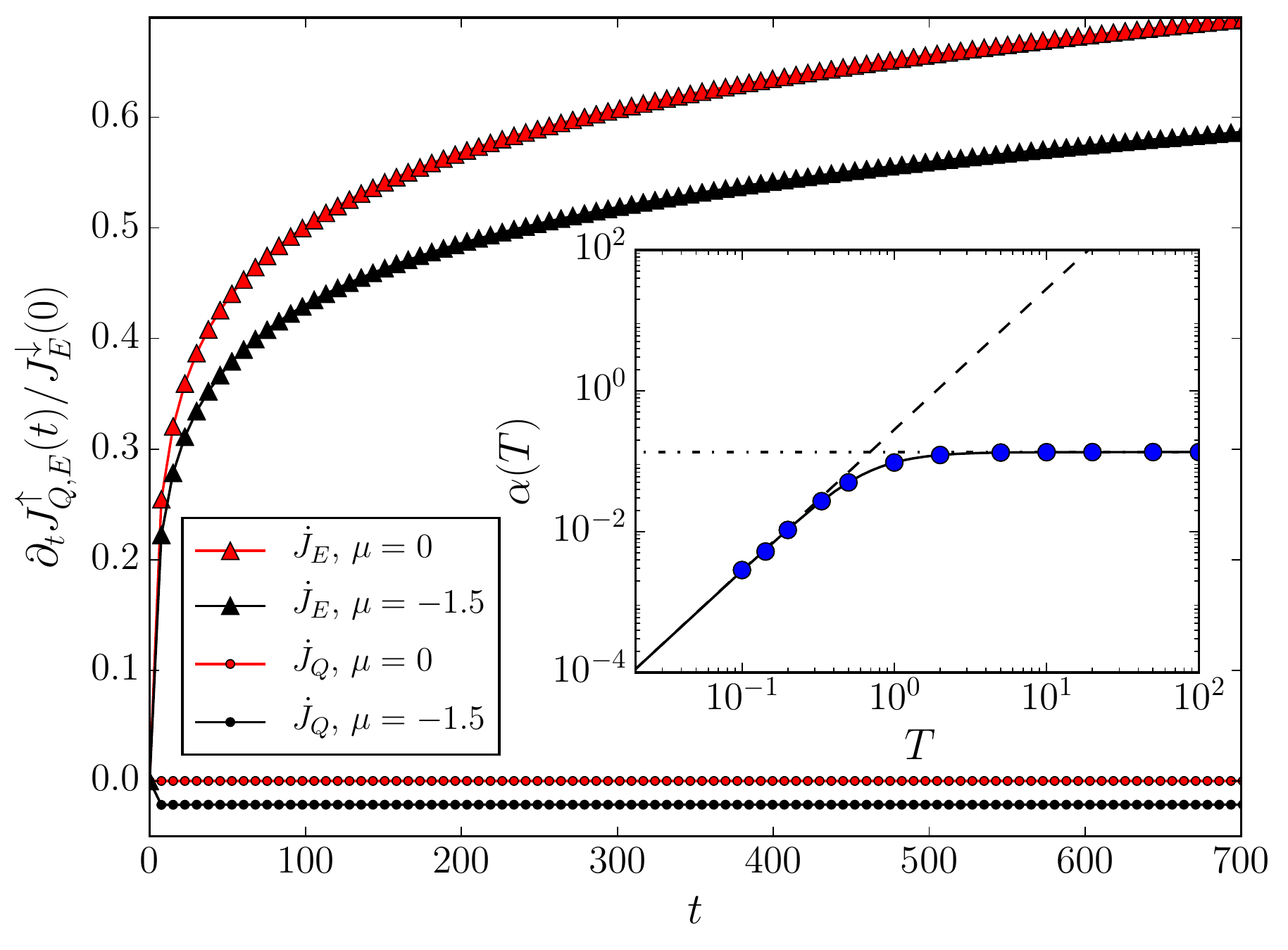}
	\caption{\label{fig:data} The growth of the heat and energy currents in the bottom layer due to the Coulomb drag, to $\mathcal O(U^2)$ in perturbation theory. At half filling $\mu=0$, no charge drag occurs due to particle-hole symmetry (red dots); this is no longer true away from half filling (black dots). In both cases, thermal drag is nonzero and the rate of change grows logarithmically in time as $\alpha(T) \log t$ (red and black triangles), rather than saturating to a constant as would be na\"ively expected. Inset: the prefactor for this log growth $\alpha(T)$ as a function of temperature. Agreement with the analyical formula of Eq.~\ref{eq:alpha} is excellent (solid line); the asymptotics are $\alpha(T) = 4 U^2 T^2 \log 2 / \pi^2$ for small $T$ (dashed line) and $\alpha(T) = 4 U^2 / 3 \pi^2$ for large $T$ (dotted line).}
\end{figure}

We have numerically checked this expression by exactly summing Eq.~\ref{eq:dtnk} on system sizes of $L>3000$ and calculating $\partial_t J_E$ and $\partial_t J$. The results are shown in Fig.~\ref{fig:data}; the logarithmic growth of the energy current is clear both at half-filling ($\mu=0$) and away from half-filling ($\mu=-1.5$). We recover the result that, as expected, there is no charge drag at half filling, confirming that thermal drag dominates in this regime, while we do notice a drag thermopower effect away from half filling. Finally, the observed dependence on temperature of the prefactor of the log, obtained by fitting at various temperatures, is in excellent agreement with Eq.~\ref{eq:alpha}, which we integrate numerically and whose asymptotics we plot. This confirms that the processes considered in this section indeed dominate the thermal drag to an excellent approximation.

A few remarks are now in order. First, the breakdown of Fermi's Golden Rule for the energy current is generic to one-dimensional systems, as any band structure will display the same kind of divergence. Second, due to the divergence, the widespread technique of linearizing the spectrum~\cite{giamarchi2004quantum} will fail badly in analyses of thermal drag. {In this case, band curvature effects may be included directly in the field theory and treated perturbatively~\cite{RevModPhys.84.1253}.} Third, the timescale for the validity of perturbation theory is parametrically reduced for thermal drag calculations: perturbation theory holds only up to a timescale $t_*^{-1} \sim U^2 \log U$. {Finally, one may consider the effects of adding a small magnetic field: to lowest order, the field would simply shift the chemical potential in the two species in opposite directions~\footnote{Excluding the effects of the field on the hopping, which are expected to be small.}, effectively breaking particle-hole symmetry. In that case, we no longer expect a vanishing charge drag. However, the logarithmic growth of the response heat current would remain, as it is present for any chemical potential, being a consequence of the band structure.}

To access longer times, we make the approximation of a linear spectrum (Luttinger liquid) and regulate the breakdown of Fermi's Golden Rule~\footnote{See the Supplemental Material, section III, where we analyze the ``generalized Golden Rule'' trick in more detail.}. Linearizing the spectrum produces a left- and a right-moving mode, described by wavevector $q_{L/R} = k \pm k_F$ with dispersion relation $E(q_{L/R}) = \mp v_F q_{L/R}$. We must then consider 8 possible scattering channels: two forward scattering channels, two Umklapp channels, and four backward scattering channels. For simplicity, we slightly modify the setup such that one spin species is kept at a temperature gradient with $k<0$ at $T_L$ and $k>0$ at $T_R$, with the other species in the ground state ($T=0$). 

Analyzing these possible scattering channels, we find that, while the Umklapp and backscattering channels give a finite rate, the forward scattering channel leads to a divergence with system size{, a one-dimensional incarnation of the well-known ``collinear scattering singularity'' in Dirac-dispersing systems~\cite{doi:10.1002/andp.201700043,PhysRevB.83.155441,RevModPhys.88.025003}}. This is due to the fact that, for the forward scattering channel, conservation of energy and momentum become the same constraint, leading to a delta function squared appearing under the scattering integral. This type of divergence was noted in Ref.~\cite{PhysRevB.78.205407} in the case of Coulomb drag for spinful Luttinger liquids. To recover a finite answer, it was proposed that one go past lowest order perturbation theory, inserting the RPA propagator in place of the bare propagator in the scattering integral (dubbed the ``generalized Fermi's golden rule''). In our case, it amounts to taking the incoming particles to have velocity $v_F$ while the outgoing particles have velocity $u$, the Luttinger velocity, which is interaction dependent. Under this prescription, we find a heat current growth rate that is actually {\it first-order} in the interaction $U$,

\begin{equation}
    \partial_t J_E^{\uparrow} \sim U \frac{2\pi^4 \log 2}{3\hbar v_F} k_B^3 (T_R^3 - T_L^3),
\end{equation}

due to the interaction-renormalized outgoing velocity cancelling a power of $U$. In sum, due to the unique divergences of heat drag as opposed to charge drag, we expect a logarithmic heat current growth rate at the shortest times that is second order in $U$, followed by a longer regime of heat current growth rate that is constant in time and first order in $U$. We emphasize that the charge drag in particle-hole symmetric systems vanishes to lowest order, and only enters at order $U^3$ (if at all); hence thermal drag is the dominant form of drag physics in this broad class of systems.

\paragraph{Long-time limit and higher dimensions.} Generally speaking, the long-time limit of this quench is outside the realm of validity of perturbation theory, and therefore inacessible. However, here we may exploit the integrability of the one-dimensional Hubbard model to make progress~\cite{esslerbook}. In particular, due to its integrability, the one-dimensional Hubbard model hosts a tower of conserved quantities, the number of which is extensive in system size. One such quantity, known as $Q_3$, differs from the total energy current operator only by a term of order $U$; that is,

\begin{align*}
J_E = t^2 \sum_{l,\sigma} i &\ (c^\dagger_{l+1,\sigma} c_{l-1,\sigma} - c^\dagger_{l-1,\sigma} c_{l+1,\sigma}) \\&- \frac{U t}{2}\sum_{l,\sigma} (j_{l-1,\sigma}+j_{l,\sigma}) (n_{l {\bar \sigma}} - 1/2),
\end{align*}

which takes the same form as $Q_3$ except for a factor of 2 in the term proportional to $U$~\cite{PhysRevLett.117.116401}. This implies that in the limit of small $U$, $J_E \approx Q_3$ and is hence conserved. (We note that even in the limit of stronger $U$, the overlap of $J_E$ with $Q_3$ will be conserved, leaving some energy current in the final state.) Under the assumption of approach to a generalized Gibbs ensemble final state~\cite{Vidmar_2016} with this same value of $Q_3$, we expect that the energy current will be equally divided between the two wires. That is,

\begin{equation}
J_E^{\uparrow}(t\to \infty) = J_E^{\downarrow}(t\to \infty) = \frac{J_E^{\downarrow}(t=0)}{2}.
\end{equation} 

The conservation of the energy current is likely a special feature due to the integrability of the Hubbard model, but we remark that in this case it leads to an intriguing hydrodynamic transport of energy current reminiscent of the Dirac fluid~\cite{Crossno1058}.

Since the source of the divergent heat drag is related to special properties of scattering in 1D, we do not expect the same divergence to appear generically for higher dimensional systems.
As a check, we have considered the Hubbard model on the square lattice with nearest-neighbor hopping~\footnote{See the Supplemental Material, section IV, where we show data for the two-dimensional Hubbard model thermal quench.}.
We have numerically explored this model for various values of the chemical potential and temperature on system sizes of up to $L_x=L_y=100$. We find that the thermal drag indeed dominates near half-filling, and it does not appear to be divergent. We defer an exhaustive analysis of the two-dimensional case to future work.

\paragraph{Discussion.} We have analyzed a thermal analogue of the Coulomb drag in interacting quantum systems with particle-hole symmetry via a quantum quench in the Hubbard model. We have found that, due to the vanishing of the charge Coulomb drag, the thermal drag effect dominates. In one dimension, its growth is drastically different than the charge drag due to the structure of the energy current operator: the short-time limit shows logarithmic non-Fermi's golden rule growth, followed by a longer regime of linear growth given by a generalized Golden rule, with the late-time limit in this case obtained from integrability arguments. 

We expect these conclusions to apply to a broad range of experimentally realizable systems, including perhaps most prominently graphene near charge neutrality. It is an interesting question whether some components of the thermal Coulomb drag may be topologically quantized in certain systems, especially in light of recent experiments on the thermal Hall effect at nonchiral Hall edges~\cite{Banerjee:2018aa}. We emphasize that, despite the vast literature on the charge Coulomb drag, the thermal drag effect is largely unexplored~\footnote{We note for completeness that another form of thermal drag was recently studied in Ref.~\cite{PhysRevB.99.201406}, but there the drag was mediated by thermal photons rather than the direct Coulomb interaction between charge carriers.}, and is ripe for further study.

\paragraph{Acknowledgements.} We thank Alex Levchenko, Fabian H. L. Essler and Thierry Giamarchi for useful conversations. WB is supported in part by the Hellman Foundation and by the DARPA DRINQS program (award D18AC00014). JEM acknowledges support from a Simons Investigatorship. WB and JEM acknowledge the hospitality of the Kavli Institute for Theoretical Physics (KITP) program ‘The Dynamics of Quantum Information’, supported in part by the National Science Foundation under Grant No. NSF PHY-1748958. T.S. acknowledges support from the Emergent Phenomena in Quantum Systems initiative of the Gordon and Betty Moore Foundation. JEM acknowledges the Center for Novel Pathways to Quantum Coherence in Materials, an Energy Frontier Research Center funded by the U.S. Department of Energy, Office of Science, Basic Energy Sciences.

\bibliography{thermal_drag}
\bibliographystyle{apsrev4-1}

\bigskip

\onecolumngrid
\newpage



\section*{Supplementary material}

# Energy drag in particle-hole symmetric systems as a quantum quench: Supplemental Material

William Berdanier,[1,*] Thomas Scaffidi,[1,2] and Joel E. Moore[1,3]

[1]*Department of Physics, University of California, Berkeley, CA 94720, USA*
[2]*Department of Physics, University of Toronto, Toronto, Ontario, M5S 1A7, Canada*
[3]*Materials Sciences Division, Lawrence Berkeley National Laboratory, Berkeley, CA 94720, USA*
(Dated: November 4, 2019)

## I. THERMAL DRAG CONDUCTIVITY KUBO FORMULA

In this appendix, we provide a derivation of the thermal drag conductivity Kubo formula presented in the main text.

Generally speaking, since there is no external perturbation (nor perturbing Hamiltonian) associated to thermal gradients and heat currents, there is no straightforward way to apply a Kubo-type formalism to heat transport. However, it is nonetheless possible to relate the presence of a temperature gradient to an "effective" perturbing Hamiltonian that would produce the same entropy growth in the system [1–3], and this leads to a Green-Kubo formula for the thermal conductivity. Here we apply this argument to the thermal drag conductivity.

Consider a bilayer system, labeled by $\sigma = 1, 2$ with unperturbed Hamiltonian $H_0$. Usually, to derive a Green-Kubo formula we would write a perturbing Hamiltonian $H_{\text{pert}} = \int dr \phi(r,t) \rho(r)$, with $\phi(r,t)$ the applied potential and $\rho(r)$ the degree of freedom of the system that couples to the potential. The time derivative is $\dot{H}_{\text{pert}} = \int dr \phi(r,t) \dot{\rho}(r)$, with $\dot{\rho}(r)$ corresponding to the unperturbed evolution of $\rho(r)$ via $i\dot{\rho} = [\rho, H_0]$ (in units of $\hbar = 1$). Now, one can show that $\dot{H}_{\text{pert}}$ balances the entropy production of the system: $\dot{H}_{\text{pert}} = -T \int dr \sigma_S$, with $\sigma_S$ the entropy source; we will use this expression to write a "perturbing Hamiltonian" even in the absence of a perturbing potential $\phi$.

Let us assume that each layer (with layer index $\sigma$) is in a situation of local equilibrium, where the local temperature is $T^\sigma(r,t) = \tilde{T}^\sigma + \delta T^\sigma(r,t)$, the local energy density is $\epsilon^\sigma(r,t) = \tilde{\epsilon}^\sigma + \delta\epsilon^\sigma(r,t)$, and the local particle density is $n^\sigma(r,t) = \tilde{n}^\sigma + \delta n^\sigma(r,t)$. Let us introduce the perturbing Hamiltonian

$$H_{\text{pert}}^\sigma = \int dr \frac{\delta T^\sigma(r,t)}{\tilde{T}^\sigma} \left( \epsilon^\sigma(r,t) - \frac{\tilde{\epsilon}^\sigma + \tilde{P}^\sigma}{\tilde{n}^\sigma} n^\sigma(r,t) \right) \tag{1}$$

where $\tilde{P}^\sigma$ is the equilibrium pressure. Appealing to the relation $TdS = dE + PdV$ and the condition $dN = 0$, we see that $TdS = d\epsilon - dn(\tilde{\epsilon} + P)/\tilde{n}$ and hence the right hand side can be interpreted as a local density of thermal energy.

Now consider the time derivative of $H_{\text{pert}}$. We assume that we are in a hydrodynamic regime $\omega\tau \ll 1$, $ql \ll 1$, with $\tau$ the typical time between collisions and $l$ the mean free path. This allows us to use hydrodynamic equations, which in linearized form are $\partial_t n(r,t) + \tilde{n} \nabla \cdot u(r,t) = 0$ and $\partial_t \epsilon(r,t) + \nabla \cdot J_E(r,t) = 0$, with $u(r,t)$ the velocity of the fluid (suppressing layer indices). The linearized energy flux is $J_E(r,t) = (\tilde{\epsilon} + \tilde{P})u(r,t) + J_Q(r,t)$. Finally, this allows us to write the time derivative of the perturbing Hamiltonian in terms of the heat current:

$$\dot{H}_{\text{pert}}^\sigma = -\int dr \frac{\delta T^\sigma(r,t)}{\tilde{T}^\sigma} \nabla \cdot J_Q^\sigma = -\tilde{T}^\sigma \int dr J_Q^\sigma \cdot \nabla \frac{1}{T^\sigma} \tag{2}$$

where in the last equality we have integrated by parts, and used the fact that, to lowest order in $\delta T$, we have $\nabla \delta T = -\tilde{T}^2 \nabla(1/T)$.

With this perturbing Hamiltonian in hand, we can now turn the crank of linear response and produce a Green-Kubo formula for the drag thermal conductivity. Write

$$\begin{aligned}\langle J_{Q,a}^{\sigma'}(r,t) \rangle &= -\frac{1}{\tilde{T}^\sigma} \int dr' \int_{-\infty}^{\infty} dt' \chi_{BA}(r-r', t-t') \nabla_b \delta T^\sigma(r',t') \\ &= -\frac{1}{\tilde{T}^\sigma} \int dr' \int_{-\infty}^{t} dt' \int_0^\beta d\lambda \langle J_{Q,b}^\sigma(r',-i\lambda) J_{Q,a}^{\sigma'}(r,t-t') \rangle \nabla_b \delta T^\sigma(r',t') \end{aligned} \tag{3}$$

with $a, b$ spatial indices, and $\chi_{BA}$ the canonical Kubo correlation function with $\dot{A}(r) = J_{Q,b}^{\sigma}$ and $B(r) = J_{Q,a}^{\sigma'}$. Fourier transforming gives the definition of the drag thermal conductivity,

$$\langle J_{Q,a}^{\sigma'}(q,\omega)\rangle = -\kappa_{ab}^{\sigma\sigma'}(q,\omega)[\nabla_b \delta T^\sigma](q,\omega). \tag{4}$$

Finally, comparison gives the Kubo formula for the drag thermal conductivity tensor,

$$\kappa_{ab}^{\sigma\sigma'}(q,\omega) = \frac{1}{VT} \lim_{\eta \to 0^+} \int_0^\infty dt e^{(i\omega-\eta)t} \int_0^\beta d\lambda \langle J_{Q,b}^\sigma(-q,-i\lambda) J_{Q,a}^{\sigma'}(q,t)\rangle, \tag{5}$$

recalling that $a, b$ are spatial indices and $\sigma, \sigma'$ are layer indices.

## II. LOGARITHMIC DIVERGENCE OF HEAT DRAG

In this appendix, we show that lowest-order time-dependent perturbation theory does not predict a Fermi's Golden Rule for the heat drag, but rather predicts a logarithmic divergence with time.

We treat charge ($a = 1$) and heat drag ($a = 2$) on the same footing. At $t = 0$, the up spins are thermal at temperature $T$, and the down spins have a distribution $n(k)$ given by thermal (at the same $T$) plus a small $\sin(ak)$ component:

$$\begin{aligned} n_{k,\downarrow}^0 &= \frac{1}{1 + e^{-2\beta_\downarrow \cos(k)}} + \eta \sin(ak) \\ n_{k,\uparrow}^0 &= \frac{1}{1 + e^{-2\beta_\uparrow \cos(k)}} \end{aligned} \tag{6}$$

The initial current is

$$J_{b,\downarrow}(t=0) = \frac{2}{N} \sum_k \sin(bk) n_{k,\downarrow}^0 = \eta \frac{1}{\pi} \int_{-\pi}^{\pi} \sin(ak)\sin(bk) = \eta \delta_{ab}. \tag{7}$$

To leading order in $U$, one finds for the time derivative of the current in the up-spins (writing $J = J_a^\uparrow$):

$$\partial_t J = \frac{2}{N} \sum_{k_3} \sin(ak_3) \partial_t \langle n_{k_3,\uparrow}(t)\rangle = -4t\eta \frac{1}{N^3} \sum_{k_1,k_2,k_3} F(k_1,k_2,k_3) |U(k_3-k_1)|^2 \operatorname{sinc}(tE), \tag{8}$$

where $F$ is a well-behaved function (no divergence), given by a sums of product of Fermi-Dirac terms and sines, and where $E = (\epsilon_{k_1} - \epsilon_{k_3} + \epsilon_{k_2} - \epsilon_{k_1+k_2-k_3})$. $k_3 - k_1$ is the net momentum transferred between the two layers, and in the Hubbard model, we have $U(q) = U$ since the interaction is on-site.

It is instructive to rewrite this as

$$\partial_t J = t \int dE\, G(E)\, \operatorname{sinc}(tE) \tag{9}$$

with

$$\begin{aligned} G(E) &= -4\eta \frac{1}{N^3} \sum_{k_1,k_2,k_3} F(k_1,k_2,k_3) |U(k_3-k_1)|^2 \delta(\epsilon_{k_1} + \epsilon_{k_2} - \epsilon_{k_3} - \epsilon_{k_1+k_2-k_3} - E) \\ &= -4\eta \frac{1}{(2\pi)^3} \int dk_1 dk_2 dk_3 F(k_1,k_2,k_3) |U(k_3-k_1)|^2 \delta(\epsilon_{k_1} + \epsilon_{k_2} - \epsilon_{k_3} - \epsilon_{k_1+k_2-k_3} - E) \\ &= -2\eta \frac{1}{(2\pi)^3} \int dk_1 dk_3 \sum_{\mu=1,2} \frac{F(k_1, k_{2,\mu}, k_3)}{|\sin(k_1 + k_{2,\mu} - k_3) - \sin(k_{2,\mu})|} |U(k_3-k_1)|^2, \end{aligned} \tag{10}$$

where $\mu$ indexes the solutions of $\epsilon_{k_1} + \epsilon_{k_2} - \epsilon_{k_3} - \epsilon_{k_1+k_2-k_3} - E = 0$, of which there are generically two for a given $k_1, k_3$. We are only interested in the $E$-even part of $G$, so let us define

$$g(E) = \frac{1}{2}(G(E) + G(-E)) \tag{11}$$

with

$$g(E) = -2\eta U^2 \frac{1}{(2\pi)^3} \int dk_1 dk_3 \sum_{\mu=1,2} \frac{f(k_1, k_{2,\mu}, k_3)}{|\sin(k_1 + k_{2,\mu} - k_3) - \sin(k_{2,\mu})|} |U(k_3 - k_1)|^2, \tag{12}$$

$$f(k_1, k_{2,\mu}, k_3) = \frac{1}{2}(F(k_1, k_{2,\mu}, k_3) + F(k_3, k_1 + k_3 - k_{2,\mu}, k_1)).$$

Writing $f$ explicitly leads to

$$2f(k_1, k_2, k_3) = \sin(ak_3)\Big[\sin(a(k_1 + k_2 - k_3))n^0_{k_3,\uparrow}(1 - n^0_{k_1,\uparrow})(1 - n^0_{k_2,\downarrow}) - \sin(ak_2)n^0_{k_3,\uparrow}n^0_{k_1+k_2-k_3,\downarrow}(1 - n^0_{k_1,\uparrow})$$
$$+ \sin(a(k_1 + k_2 - k_3))(1 - n^0_{k_3,\uparrow})n^0_{k_1,\uparrow}n^0_{k_2,\downarrow} - \sin(ak_2)(1 - n^0_{k_3,\uparrow})(1 - n^0_{k_1+k_2-k_3,\downarrow})n^0_{k_1,\uparrow}\Big]$$
$$+ \sin(ak_1)\Big[\sin(a(k_2))n^0_{k_1,\uparrow}(1 - n^0_{k_3,\uparrow})(1 - n^0_{k_1+k_2-k_3,\downarrow}) - \sin(a(k_1 + k_2 - k_3))n^0_{k_1,\uparrow}n^0_{k_2,\downarrow}(1 - n^0_{k_3,\uparrow})$$
$$+ \sin(ak_2)(1 - n^0_{k_1,\uparrow})n^0_{k_3,\uparrow}n^0_{k_1+k_2-k_3,\downarrow} - \sin(a(k_1 + k_2 - k_3))(1 - n^0_{k_1,\uparrow})(1 - n^0_{k_2,\downarrow})n^0_{k_3,\uparrow}\Big], \tag{13}$$

which can be rearranged as

$$2f(k_1, k_2, k_3) = \Big(\sin(ak_3) - \sin(ak_1)\Big)\Big[-\sin(ak_2)\Big\{(1 - n^0_{k_3,\uparrow})(1 - n^0_{k_1+k_2-k_3,\downarrow})n^0_{k_1,\uparrow} + n^0_{k_3,\uparrow}n^0_{k_1+k_2-k_3,\downarrow}(1 - n^0_{k_1,\uparrow})\Big\}$$
$$+ \sin(a(k_1 + k_2 - k_3))\Big\{(1 - n^0_{k_3,\uparrow})n^0_{k_1,\uparrow}n^0_{k_2,\downarrow} + n^0_{k_3,\uparrow}(1 - n^0_{k_1,\uparrow})(1 - n^0_{k_2,\downarrow})\Big\}\Big]. \tag{14}$$

Clearly, the only source of divergence is the factor $\frac{1}{|\sin(k_1+k_{2,\mu}-k_3)-\sin(k_{2,\mu})|}$. This term diverges in two cases, $k_1 = k_3$, and $k_1 + k_2 - k_3 = \pi - k_2$. In the case of $k_1 = k_3$, one can see that $f$ always vanishes, and the divergence is therefore cured.

The case of $k_1 + k_2 - k_3 = \pi - k_2$ is more tricky. Plugging this in $f$ leads to

$$2f_{\text{div}}(k_1, k_2, k_3) = \Big(\sin(ak_3) - \sin(ak_1)\Big)\Big[n^0_{k_3,\uparrow}(1 - n^0_{k_1,\uparrow})(1 - n^0_{k_2,\downarrow})\sin(a(\pi - k_2)) - n^0_{k_3,\uparrow}n^0_{\pi-k_2,\downarrow}(1 - n^0_{k_1,\uparrow})\sin(ak_2)$$
$$+ (1 - n^0_{k_3,\uparrow})n^0_{k_1,\uparrow}n^0_{k_2,\downarrow}\sin(a(\pi - k_2)) - (1 - n^0_{k_3,\uparrow})(1 - n^0_{\pi-k_2,\downarrow})n^0_{k_1,\uparrow}\sin(ak_2)\Big]. \tag{15}$$

Focusing on half-filling, one has the property that $(1 - n^0_k) = n^0_{\pi-k}$. This leads to

$$2f_{\text{div}}(k_1, k_2, k_3) = \Big(\sin(a(\pi - k_2)) - \sin(ak_2)\Big)\Big(\sin(ak_3) - \sin(ak_1)\Big)\Big[n^0_{k_3,\uparrow}(1 - n^0_{k_1,\uparrow})(1 - n^0_{k_2,\downarrow}) + (1 - n^0_{k_3,\uparrow})n^0_{k_1,\uparrow}n^0_{k_2,\downarrow}\Big]. \tag{16}$$

The first factor vanishes for odd $a$, but is finite for even $a$, and will remain finite unless we fine-tune $|U(k_3 - k_1)|^2$. This is why charge drag is not divergent (and actually zero at half-filling, but could be finite away from half-filling), while heat drag is divergent.

### A. Taking care of the divergence at $k_1 + k_2 - k_3 = \pi - k_2$

Let us first try naively at $E = 0$. In that case, the solutions for $k_2$ are $k_{2,\nu=1} = \pi - k_1$ and $k_{2,\nu=2} = k_3$. Plugging this in $k_1 + k_2 - k_3 = \pi - k_2$ leads to a line of divergences at $k_1 + k_3 = \pi$, in both cases.

At small but finite $E$, we can regularize the $\frac{1}{|\sin(k_1+k_{2,\mu}-k_3)-\sin(k_{2,\mu})|}$ factor as

$$\frac{1}{|\sin(k_1 + k_{2,\mu} - k_3) - \sin(k_{2,\mu})|} \mapsto \text{Re}\left[\frac{1}{\sqrt{(\sin(k_3) - \sin(k_1))^2 - 2E(\cos(k_3) - \cos(k_1))}}\right]. \tag{17}$$

We finally find

$$g(E) = -\frac{2\eta}{(2\pi)^3} \int dk_1 dk_3 |U(k_3 - k_1)|^2 \sum_{\mu=1,2} f(k_1, k_{2,\mu}, k_3) \text{Re}\left[\frac{1}{\sqrt{(\sin(k_3) - \sin(k_1))^2 - 2E(\cos(k_3) - \cos(k_1))}}\right]. \tag{18}$$

We change variables first, using $k_\pm = k_1 \pm k_3$, obtaining

$$g(E) = -\frac{\eta}{(2\pi)^3} \int dk_+ dk_- |U(-k_-)|^2 \sum_{\mu=1,2} f_\mu(k_+, k_-) \text{Re}\left[\frac{1}{\sqrt{4\cos(k_+/2)^2 \sin(k_-/2)^2 - 4E\sin(k_+/2)\sin(k_-/2)}}\right]. \tag{19}$$

On physical grounds, for time-reversal symmetric interactions, we expect $U(-q) = U(q)$, so from now on let us assume that $U$ is even. Since the integral will be dominated by the near-divergence of the denominator close to $k_+ = \pi$, we can approximate $f$ to take its value on that line, which is the same for the two solutions:

$$g(E) = -\frac{2\eta U^2}{(2\pi)^3} \int dk_+ dk_- \, |U(k_-)|^2 \, f(k_+ = \pi, k_-) \text{Re}\left[\frac{1}{\sqrt{4\cos(k_+/2)^2 \sin(k_-/2)^2 - 4E\sin(k_+/2)\sin(k_-/2)}}\right]. \tag{20}$$

We can now perform the integral over $k_+$. Writing $k_+ = \pi + \epsilon$, one finds

$$g(E) = -\frac{4\eta}{(2\pi)^3} \int_{-\pi}^{\pi} dk_- \, |U(k_-)|^2 \int_{-K}^{+K} d\epsilon f(k_+ = \pi, k_-) \text{Re}\left[\frac{1}{\sqrt{\epsilon^2 \sin(k_-/2)^2 - 4E\sin(k_-/2)}}\right] \tag{21}$$

where we added a factor of 2 because we restricted $k_1$ and $k_3$ to lie in the first quadrant (which leads to $k_-$ running from $-\pi$ to $\pi$). $K$ is a large momentum cutoff, which will lead to a non-divergent piece that will be discarded.

Using

$$\int_{-K}^{+K} dx \text{Re}\left[\frac{1}{\sqrt{x^2 - b}}\right] \to -\log|b| \tag{22}$$

in the small $b$ limit, one finds

$$g(E) = \frac{4\eta U^2}{(2\pi)^3} \int_{-\pi}^{\pi} dk_- \, |U(k_-)|^2 \, \frac{f(k_+ = \pi, k_-)}{|\sin(k_-/2)|} \log\left(\frac{4|E|}{|\sin(k_-/2)|}\right). \tag{23}$$

Focusing on the divergent piece, we finally find

$$g(E) = \alpha \log|E|, \qquad \alpha = 4\eta U^2 \frac{1}{(2\pi)^3} \int_{-\pi}^{\pi} dk_- \, |U(k_-)|^2 \, \frac{f(k_+ = \pi, k_-)}{|\sin(k_-/2)|}, \tag{24}$$

where, at half-filling and $k_+ = \pi$, one has

$$2f(k_1, k_2, k_3) = \Big(\sin(a(\pi - k_2)) - \sin(ak_2)\Big)\Big(\sin(ak_3) - \sin(ak_1)\Big)\Big[n_{k_3,\uparrow}^0 (1 - n_{k_1,\uparrow}^0)(1 - n_{k_2,\downarrow}^0) + (1 - n_{k_3,\uparrow}^0) n_{k_1,\uparrow}^0 n_{k_2,\downarrow}^0\Big]. \tag{25}$$

Focusing on heat drag ($a = 2$), one finds

$$2f(k_1, k_3) = -4\sin(k_-)^2 \Big((n_{k_3,\uparrow}^0)^2 n_{k_1,\downarrow}^0 + (n_{k_1,\uparrow}^0)^2 n_{k_3,\downarrow}^0\Big) = -4\sin(k_-)^2 \Big((n_\uparrow(\epsilon)^2 n_\downarrow(-\epsilon) + (n_\uparrow(-\epsilon)^2 n_\downarrow(\epsilon)\Big), \tag{26}$$

where $\epsilon(k_-) = -2\sin(k_-/2)$ and $n_{\uparrow,\downarrow}(\epsilon) = 1/(1 + e^{\beta_{\uparrow,\downarrow}\epsilon})$.

Let us focus on $T_\uparrow = T_\downarrow$ for now, leading to

$$2f(k_1, k_3) = -4\sin(k_-)^2 \Big((n(\epsilon)^2 n(-\epsilon) + (n(-\epsilon)^2 n(\epsilon)\Big) = -4\sin(k_-)^2 n(\epsilon) n(-\epsilon). \tag{27}$$

We finally do the integral over $E$, by using

$$\int_{-\infty}^{\infty} dx \log(x) \, \text{sinc}(xt) = -\pi \frac{\gamma + \log(t)}{t}, \tag{28}$$

where $\gamma$ is the Euler-Mascheroni constant. Keeping only the dominant term, this leads to

$$\partial_t J = t\alpha \int dE \, \log(E) \, \text{sinc}(tE) = \alpha' \log(t) + \mathcal{O}(1), \tag{29}$$

where

$$\alpha' = -\pi\alpha = \frac{\eta U^2}{\pi^2} \int_{-\pi}^{\pi} dk_- \, |U(k_-)|^2 \, \frac{\sin(k_-)^2}{|\sin(k_-/2)|} n(2\sin(k_-/2)) n(-2\sin(k_-/2)) \equiv \frac{\eta U^2}{\pi^2} I(T), \tag{30}$$

where we write the interaction has a having a characteristic scale $U$ such that $U(q) = Uu(q)$, and

$$I(T) \equiv \int_{-\pi}^{\pi} dk_- \, |u(k_-)|^2 \, \frac{\sin(k_-)^2}{|\sin(k_-/2)|} n(2\sin(k_-/2)) n(-2\sin(k_-/2)). \tag{31}$$

Normalizing by the initial current in the down spins leads to

$$\frac{\partial_t J_\uparrow}{J_\downarrow(t=0)} = \frac{U^2}{\pi^2} I(T) \log(t) \tag{32}$$

as shown in the main text. Finally, we can massage the expression for $I(T)$, using the fact that the integrand is even, to obtain

$$I(T) = 2 \int_0^\pi dk \, |u(k)|^2 \frac{\sin^2 k}{\sin(k/2)} \frac{1}{1+\exp(-2t\beta \sin(k/2))} \frac{1}{1+\exp(2t\beta \sin(k/2))} \tag{33}$$

$$= \int_0^\pi dk \, |u(k)|^2 \frac{\sin^2(k) \csc(k/2)}{1+\cosh(2t\beta \sin(k/2))}, \tag{34}$$

where $t$ is the Hubbard hopping parameter (not time). For the particular case $U(k) = U$, we have

$$I(T) = \int_0^\pi dk \frac{\sin^2(k) \csc(k/2)}{1+\cosh(2t\beta \sin(k/2))}. \tag{35}$$

We can analyze its limits as follows: for small $\beta \ll 1$, the denominator of the integrand is approximately constant, so

$$I(T) \approx \frac{1}{2} \int_0^\pi dk \sin^2(k) \csc(k/2) = 4/3. \tag{36}$$

The other limit of $\beta \gg 1$ requires a bit more work, but the key is that the denominator is a sharply peaked function. Rewriting in terms of $\epsilon = 2t \sin(k/2)$ with $t = 1$ for convenience,

$$I(T) = \int_{-2}^2 d\epsilon \frac{|\sin(2 \arcsin(\epsilon/2))|}{1+\cosh(\beta \epsilon)}. \tag{37}$$

Noting that the integrand is strongly peaked around $\epsilon = 0$ when $\beta \gg 1$, we can extend the limits of integration and Taylor expand the numerator in $\epsilon$, yielding

$$I(T) \approx \int_{-\infty}^\infty d\epsilon \frac{|\epsilon|}{1+\cosh(\beta \epsilon)}. \tag{38}$$

Finally, we change variables again to $\epsilon' = \beta \epsilon$, yielding

$$I(T) \approx \frac{2}{\beta^2} \int_0^\infty d\epsilon' \frac{\epsilon'}{1+\cosh(\epsilon')} = \frac{4 \ln 2}{\beta^2}. \tag{39}$$

A similar analysis may be made for other forms of the interaction $U(q)$.

### B. Validity of perturbation theory

Perturbation theory relies on the occupation numbers (and therefore the currents) being close to their initial values, i.e. $|n_k(t) - n_k(0)| \ll 1$.

The current at time $t$ is given by the integral of the rate,

$$\frac{J_\uparrow(t)}{J_\downarrow(t=0)} = \int_0^t dt' \frac{\partial_t J_\uparrow}{J_\downarrow(t=0)} = 2\frac{U^2}{\pi^2} I(T) \, t \log(t) + \mathcal{O}(t) \tag{40}$$

where we have neglected an $\mathcal{O}(1)$ term in the rate, which leads to an $\mathcal{O}(t)$ term in the current.

Perturbation theory should be accurate as long as

$$\frac{J_\uparrow(t)}{J_\downarrow(t=0)} \ll 1, \tag{41}$$

which leads to

$$t \log(t) \ll \frac{\pi^2}{2U^2 I(T)}. \tag{42}$$

To a good approximation, this is equivalent to

$$t \ll t^* \sim \frac{1}{U^2 \log |U|}. \tag{43}$$

Remarkably, this is parametrically shorter than the usual time for FGR to break, which is $\sim 1/U^2$.

### III. GENERALIZED FERMI'S GOLDEN RULE

In this appendix, we demonstrate that Fermi's Golden Rule applied to thermal drag in the spinful Luttinger liquid produces a divergence, and detail a method to correct this behavior known as the "generalized Golden Rule" [4].

Let us do an explicit calculation of $J_E(t)$ at short times and small $U$ using Fermi's Golden Rule. We set up the problem as follows. Consider the one-dimensional Hubbard model. Prepare the spin-up channel at temperature $T$, such that we have the initial distribution function

$$n^0_\uparrow(k) = f_T(k) \tag{44}$$

where we define

$$f_T(k) = \frac{1}{1 + \exp(-\frac{1}{k_B T} 2 t_h \cos k)} \tag{45}$$

letting the hopping parameter be $t_h$ to avoid confusion with time $t$. Prepare the spin-down channel with a bath at temperature $T_L$ for the left-movers, $T_R$ for the right-movers (so the left side is at $T_R$ and the right side at $T_L$). Then the distribution function is

$$n^0_\downarrow(k) = \begin{cases} f_{T_L}(k) & k < 0 \\ f_{T_R}(k) & k > 0 \end{cases} \tag{46}$$

which should be normalized already since the Fermi-Dirac distribution at half filling is symmetric about $k = 0$. Note that it is discontinuous at $k = 0$, though. Now, the Fermi's Golden Rule transition rate between the states $|k_1 \uparrow, k_2 \downarrow\rangle \to |k_3 \uparrow, k_4 \downarrow\rangle$ is [5]

$$Q^{k_3\uparrow,k_4\downarrow}_{k_1\uparrow,k_2\downarrow} = \frac{2\pi}{\hbar} \left|\langle k_1 \uparrow, k_2 \downarrow |\hat{U}| k_3 \uparrow, k_4 \downarrow\rangle\right|^2 \delta(E_{k_1} + E_{k_2} - E_{k_3} - E_{k_4}) = \frac{2\pi}{\hbar} U^2 \delta(k_1 + k_2 - k_3 - k_4) \delta(E_{k_1} + E_{k_2} - E_{k_3} - E_{k_4}) \tag{47}$$

which explicitly conserves momentum and energy. Since the scattering is reversible, $Q^{k_3\uparrow,k_4\downarrow}_{k_1\uparrow,k_2\downarrow} = Q^{k_1\uparrow,k_2\downarrow}_{k_3\uparrow,k_4\downarrow}$. Hence the rate of change of $n^\uparrow_{k_1}$ is

$$\begin{aligned}\partial_t n^\uparrow_{k_1} &= \int dk_2 dk_3 dk_4 [n^\uparrow_{k_3} n^\downarrow_{k_4}(1-n^\uparrow_{k_1})(1-n^\downarrow_{k_2}) - n^\uparrow_{k_1} n^\downarrow_{k_2}(1-n^\uparrow_{k_3})(1-n^\downarrow_{k_4})] Q^{k_3\uparrow,k_4\downarrow}_{k_1\uparrow,k_2\downarrow} \\ &= \frac{2\pi}{\hbar} U^2 \int dk_2 dk_3 dk_4 [n^\uparrow_{k_3} n^\downarrow_{k_4}(1-n^\uparrow_{k_1})(1-n^\downarrow_{k_2}) - n^\uparrow_{k_1} n^\downarrow_{k_2}(1-n^\uparrow_{k_3})(1-n^\downarrow_{k_4})] \\ &\quad \times \delta(k_1 + k_2 - k_3 - k_4) \delta(E_{k_1} + E_{k_2} - E_{k_3} - E_{k_4})\end{aligned} \tag{48}$$

with $E_k = -2t_h \cos k$. Now let's linearize the spectrum. Write

$$\begin{cases} k = k_F + q_R & k > 0 \\ k = -k_F + q_L & k < 0 \end{cases} \tag{49}$$

TABLE I. Scattering processes in FGR calculation.

| 1↑ | 2↓ | 3↑ | 4↓ | process |
|---|---|---|---|---|
| L | L | L | L | forward |
| R | R | R | R | |
| R | R | L | L | Umklapp |
| L | L | R | R | |
| R | L | R | L | backscattering |
| R | L | L | R | |
| L | R | R | L | |
| L | R | L | R | |

and approximate cosine spectrum by a pair of lines:

$$\begin{cases} E_k = E_{k_F} + v_F q^R & k > 0 \\ E_k = E_{-k_F} - v_F q^L & k < 0 \end{cases} \tag{50}$$

We take the left and right movers $q^{R,L}$ to be defined in some bandwidth about the Fermi wavevector: $q \in [-\Lambda, \Lambda]$, with $k$ outside this range not contributing to the low-energy physics. We then take the limit $\Lambda \to \infty$, assuming that large $q$ does not contribute very much.

At half-filling, $E_{k_F} = E_{-k_F} = 0$, $k_F = \pi/2a$ (where the lattice spacing $a = 1$) and $v_F = 2t_h$. Hence $E_k = \pm v_F q^{R,L}$. This means that

$$\int_{BZ} dk = \int_{k<0} dk + \int_{k>0} dk = \int_{-\infty}^{\infty} dq^L + \int_{-\infty}^{\infty} dq^R \tag{51}$$

hence

$$\int dk_2 dk_3 dk_4 f(n_{k_i}) \delta(\sum k) \delta(\sum E_k) = (\int dq_2^L + \int dq_2^R)(\int dq_3^L + \int dq_3^R)(\int dq_4^L + \int dq_4^R) f(n_{k_i}) \delta(\sum k) \delta(\sum E_k) \tag{52}$$

so we've now turned one integral into 8 integrals (and we integrate over $q_1^{R,L}$ as allowed by momentum conservation). The following processes then contribute to $\partial_t n_k^\uparrow$:

### A. Forward scattering

Let's look at just one of these, say the LLLL channel. Recalling that the energy current is

$$\partial_t J_E^\uparrow = \int dk_1 v_{k_1} E_{k_1} \partial_t n_{k_1}^\uparrow \tag{53}$$

with $v_{k_1} = -v_F$ and $E_{k_1} = -v_F q_1^L$, the LLLL channel contribution to $\partial_t J_E^\uparrow$ is

$$\frac{2\pi}{\hbar} U^2 \int dq_1^L dq_2^L dq_3^L dq_4^L \delta(q_3^L + q_4^L - q_2^L - q_1^L) \delta(v_F(q_3^L + q_4^L - q_2^L - q_1^L)) \\ \times [f_T(q_3^L) f_{T_L}(q_4^L)(1 - f_T(q_1^L))(1 - f_{T_L}(q_2^L)) - f_T(q_1^L) f_{T_L}(q_2^L)(1 - f_T(q_3^L))(1 - f_{T_L}(q_4^L))] v_F^2 q_1^L. \tag{54}$$

Using the delta function identity $\delta(ax) = \delta(x)/|a|$ to pull out a factor $1/v_F$, and integrating $dq_2^L$, such that $q_2^L = q_3^L + q_4^L - q_1^L$, we get

$$\frac{2\pi}{\hbar} U^2 v_F \delta(0) \int dq_1^L dq_3^L dq_4^L [f_T(q_3^L) f_{T_L}(q_4^L)(1 - f_T(q_1^L))(1 - f_{T_L}(q_3^L + q_4^L - q_1^L)) \\ - f_T(q_1^L) f_{T_L}(q_3^L + q_4^L - q_1^L)(1 - f_T(q_3^L))(1 - f_{T_L}(q_4^L))] q_1^L. \tag{55}$$

This $\delta(0)$ factor we can take to be regularized as $\delta(0) = L$; it comes from conservation of momentum and energy being the same constraint for a linear spectrum. We can also now use the Fermi-Dirac distribution identity $1 - f_{T_L}(q_3^L + q_4^L - q_1^L) = f_{T_L}(q_1^L - q_3^L - q_4^L)$.

Now let's approximate $T = 0$ so that we only have one temperature scale ($T_L$) in the problem. Then $f_T(q^L) = \Theta(-v_F q^L)$, where $\Theta$ is the Heaviside step function,

$$\Theta(\epsilon) = \begin{cases} 0 & \epsilon < 0 \\ 1 & \epsilon > 0 \end{cases}.$$

We use this $\Theta$ function to fix the limits of integration. Plugging in the FD distributions:

$$\frac{2\pi}{\hbar} U^2 v_F L \left\{ \int_{-\infty}^0 dq_1^L \int_0^\infty dq_3^L \int_{-\infty}^\infty dq_4^L \frac{q_1^L}{1+\exp(-\beta_L v_F q_4^L)} \frac{1}{1+\exp(\beta_L v_F (q_4^L - (q_1^L - q_3^L)))} \right. \tag{56}$$
$$\left. - \int_0^\infty dq_1^L \int_{-\infty}^0 dq_3^L \int_{-\infty}^\infty dq_4^L \frac{q_1^L}{1+\exp(\beta_L v_F q_4^L)} \frac{1}{1+\exp(-\beta_L v_F (q_4^L - (q_1^L - q_3^L)))} \right\}.$$

Focusing on the first integral,

$$\int_{-\infty}^0 dq_1^L \int_0^\infty dq_3^L \int_{-\infty}^\infty dq_4^L \frac{q_1^L}{1+\exp(-\beta_L v_F q_4^L)} \frac{1}{1+\exp(\beta_L v_F (q_4^L - (q_1^L - q_3^L)))}$$
$$= \int_{-\infty}^0 dq_1^L \int_0^\infty dq_3^L \frac{1}{\beta_L v_F} \frac{q_1^L (q_3^L - q_1^L)}{1 - \exp(\beta_L v_F (q_3^L - q_1^L))} \tag{57}$$
$$= \int_{-\infty}^0 dq_1^L \left[ \frac{1}{(\beta_L v_F)^2} \text{Li}_2(e^{\beta_L v_F q_1^L}) + \frac{1}{\beta_L v_F} q_1^L \log(1 - e^{\beta_L v_F q_1^L}) \right] q_1^L$$
$$= -\frac{\pi^4}{30} \frac{1}{(\beta_L v_F)^4}$$

with $\text{Li}_2(z)$ the (order 2) polylogarithm function. Similarly the second integral contributes

$$-\int_0^\infty dq_1^L \int_{-\infty}^0 dq_3^L \int_{-\infty}^\infty dq_4^L \frac{q_1^L}{1+\exp(\beta_L v_F q_4^L)} \frac{1}{1+\exp(-\beta_L v_F (q_4^L - (q_1^L - q_3^L)))} = -\frac{\pi^4}{30} \frac{1}{(\beta_L v_F)^4}. \tag{58}$$

Thus, the LLLL channel contributes

$$-\frac{2\pi}{\hbar} U^2 v_F L \frac{\pi^4}{15} \frac{1}{(\beta_L v_F)^4} \tag{59}$$

to $\partial_t J_E(t=0)$. Now, by the same logic the RRRR channel contributes

$$+\frac{2\pi}{\hbar} U^2 v_F L \frac{\pi^4}{15} \frac{1}{(\beta_R v_F)^4}. \tag{60}$$

Hence the total contribution from forward scattering is, using $v_F = 2t_h$ at half filling,

$$\frac{\pi^5}{60\hbar} \frac{U^2}{t_h^3} L k_B^4 \left( T_R^4 - T_L^4 \right). \tag{61}$$

For ease of reference, $\pi^5/60 \approx 5.1$.

### B. Umklapp

Now consider the Umklapp channels, RRLL and LLRR. First focus on RRLL. Conservation of momentum reads $\sum k = (k_F + q_1) + (k_F + q_2) - (-k_F + q_3) - (-k_F + q_4) = 4k_F + q_1 + q_2 - q_3 - q_4 = 0$. Since we are at half filling,

$4k_F = 4(\pi/2) = 2\pi = 0 \mod 2\pi$. Hence conservation of momentum is simply $q_1 + q_2 - q_3 - q_4 = 0$. Conservation of energy reads $\sum E_k = 0 = v_F q_1 + v_F q_2 - (-v_F q_3 - v_F q_4) = v_F(q_1 + q_2 + q_3 + q_4)$. Hence, the contribution to $\partial_t J_E$ is

$$\frac{2\pi}{\hbar} v_F^2 U^2 \left\{ \int_{-\infty}^{0} dq_1 \int_{-\infty}^{\infty} dq_2 \int_{-\infty}^{0} dq_3 \int_{-\infty}^{\infty} dq_4 f_{T_R}(-q_2) f_{T_L}(q_4) \delta(q_1 + q_2 - q_3 - q_4) \delta(v_F(q_1 + q_2 + q_3 + q_4)) \right. \\ \left. - \int_{0}^{\infty} dq_1 \int_{-\infty}^{\infty} dq_2 \int_{0}^{\infty} dq_3 \int_{-\infty}^{\infty} dq_4 f_{T_R}(q_2) f_{T_L}(-q_4) \delta(q_1 + q_2 - q_3 - q_4) \delta(v_F(q_1 + q_2 + q_3 + q_4)) \right\}. \tag{62}$$

The integral over $q_2$ sets $q_2 = q_3 + q_4 - q_1$, and then conservation of energy reads

$$\delta(v_F(q_1 + (q_3 + q_4 - q_1) + q_3 + q_4)) = \frac{1}{2v_F} \delta(q_3 + q_4).$$

Integrating over $q_4$ then yields

$$\frac{\pi}{\hbar v_F} U^2 \left\{ \int_0^\infty dq_1 q_1 \int_0^\infty dq_3 f_{T_R}(q_1) f_{T_L}(-q_3) - \int_{-\infty}^0 dq_1 q_1 \int_{-\infty}^0 dq_3 f_{T_R}(-q_1) f_{T_L}(q_3) \right\} = \frac{\pi^3 \log 2}{6\hbar} \frac{U^2}{v_F^2} \frac{1}{\beta_R^2 \beta_L}. \tag{63}$$

Similarly, LLRR yields $-\frac{\pi^3 \log 2}{6\hbar} \frac{U^2}{v_F^2} \frac{1}{\beta_L^2 \beta_R}$. Hence, using $v_F = 2t_h$, the total contribution from Umklapp scattering is

$$\frac{\pi^3 \log 2}{24\hbar} \frac{U^2}{t_h^2} k_B^3 \left( T_R^2 T_L - T_L^2 T_R \right). \tag{64}$$

### C. Backscattering

Finally, consider the four backscattering channels: RLRL, LRLR, RLLR, LRRL. Taking RLRL, the integrals we want to compute are

$$v_F^2 \frac{2\pi}{\hbar} U^2 \left\{ \int_0^\infty dq_1 q_1 \int_{-\infty}^\infty dq_2 \int_{-\infty}^0 dq_3 \int_{-\infty}^\infty dq_4 f_{T_L}(-q_2) f_{T_L}(q_4) \delta(\sum k) \delta(\sum E_k) \right. \\ \left. - \int_{-\infty}^0 dq_1 q_1 \int_{-\infty}^\infty dq_2 \int_0^\infty dq_3 \int_{-\infty}^\infty dq_4 f_{T_L}(q_2) f_{T_L}(-q_4) \delta(\sum k) \delta(\sum E_k) \right\} \tag{65}$$

with $\sum k = q_1 + q_2 - q_3 - q_4$ and $\sum E_k = v_F(q_1 - q_3 - q_2 + q_4)$. Then when we integrate $dq_2$, conservation of momentum will set $q_2 = q_3 + q_4 - q_1$ and hence the conservation of energy delta function will read

$$\delta(2v_F(q_1 - q_3)) = \frac{1}{2v_F} \delta(q_1 - q_3)$$

and will pick out $q_1 = q_3$. However, $q_1 = q_3$ means both are 0 in the above integrals! This picks out just one point under the integral sign (i.e. a set of measure 0), so the contribution from this process *vanishes*. Similarly, LRLR vanishes. This is intuitively reasonable, since if you draw out the setup in RLRL and LRLR, the only way to conserve energy and momentum is for the particles to not scatter at all.

Now consider the true backscattering processes, LRRL and RLLR. Focusing on LRRL:

$$v_F^2 \frac{2\pi}{\hbar} U^2 \left\{ \int_{-\infty}^0 dq_1 q_1 \int_{-\infty}^\infty dq_2 \int_{-\infty}^0 dq_3 \int_{-\infty}^\infty dq_4 f_{T_R}(-q_2) f_{T_L}(q_4) \delta(\sum k) \delta(\sum E_k) \right. \\ \left. - \int_0^\infty dq_1 q_1 \int_{-\infty}^\infty dq_2 \int_0^\infty dq_3 \int_{-\infty}^\infty dq_4 f_{T_R}(q_2) f_{T_L}(-q_4) \delta(\sum k) \delta(\sum E_k) \right\} \tag{66}$$

with $\sum k = q_1 + q_2 - q_3 - q_4$ and $\sum E_k = v_F(q_2 - q_1 + q_4 - q_3)$. Performing the integral over $q_2$ sets $q_2 = q_3 + q_4 - q_1$ hence conservation of energy becomes

$$\delta(2v_F(q_4 - q_1)) = \frac{1}{2v_F}\delta(q_4 - q_1).$$

Then performing the integral over $q_4$ gives

$$\frac{\pi v_F}{\hbar}U^2\left\{\int_{-\infty}^0 dq_1 q_1 \int_{-\infty}^0 dq_3 f_{T_R}(-q_3)f_{T_L}(q_1) - \int_0^\infty dq_1 q_1 \int_0^\infty dq_3 f_{T_R}(q_3)f_{T_L}(-q_1)\right\} = -\frac{\pi^3 \log 2}{6\hbar}U^2 v_F^2 \frac{1}{\beta_L^2 \beta_R}. \tag{67}$$

The total contribution from from backscattering is then

$$\frac{\pi^3 \log 2}{24\hbar}\frac{U^2}{t_h^2}k_B^3\left(T_R^2 T_L - T_L^2 T_R\right). \tag{68}$$

For ease of reference, $\frac{\pi^3 \log 2}{24} \approx 0.9$.

### D. "Generalized Golden Rule" trick and foward scattering

We found that the forward scattering process leads to a divergence in the case of the (spinful) Hubbard model, and generally between two wires. As described in [4], the forward scattering integral is finite with spinless fermions but diverges when we have multiple spin species. They use a trick (referred to as the 'generalized Fermi's golden rule') to fix this divergence, which we adapt here.

The issue with the forward scattering channel is that conservation of momentum and conservation of energy are the same constraint due to the linear spectrum, so we get a delta function squared under the integral. The intuitive picture for why it is inconsistent to use $v_F$ for the energy conservation constraint in a Fermi's golden rule calculation is that the Fermi velocity is renormalized by interactions at first order in $U$. The patch that Yashenkin et al. prescribe is to replace the usual energy constraint $v_F q_{in} = v_F q_{out}$ with $v_F q_{in} = u q_{out}$, with $u$ the Luttinger liquid velocity. That is, we replace

$$\delta(q_1 + q_2 - q_3 - q_4)\delta(v_F(q_1 + q_2) - v_F(q_3 + q_4)) \mapsto \delta(q_1 + q_2 - q_3 - q_4)\delta(v_F(q_1 + q_2) - u(q_3 + q_4)), \tag{69}$$

with Luttinger velocity (using $g_4 = g_2 = U$)

$$u = v_F\sqrt{1 + U/\pi v_F}. \tag{70}$$

For completeness, the $K$ parameter is $K = 1/\sqrt{1 + U/\pi v_F}$ (note that for Hubbard, since all $g$ are equal we get $uK = v_F$). Now, using this new delta function, the LLLL contribution to $\partial_t J_E(t=0)$ is

$$\frac{2\pi}{\hbar}U^2 v_F^2 \int_{\mathbb{R}^4} dq_{1234}\delta(q_1 + q_2 - q_3 - q_4)\delta(v_F(q_1 + q_2) - u(q_3 + q_4)) \\ \times q_1\{f_T(q_3)f_{T_L}(q_4)f_T(-q_1)f_{T_L}(-q_2) - f_T(q_1)f_{T_L}(q_2)f_T(-q_3)f_{T_L}(-q_4)\} \tag{71}$$

using the fact that $f_T(-k) = 1 - f_T(k)$ for the Fermi-Dirac distribution. Integrate over $q_2$, which enforces the first delta function, setting $q_2 = q_3 + q_4 - q_1$, and setting the second delta function to $\delta((u - v_F)(q_3 + q_4)) = \delta(q_3 + q_4)/|u - v_F|$. We get

$$\frac{2\pi}{\hbar}U^2\frac{v_F^2}{|u - v_F|}\int_{\mathbb{R}^3} dq_{134}\delta(q_3 + q_4) \\ \times q_1\{f_T(q_3)f_{T_L}(q_4)f_T(-q_1)f_{T_L}(-(q_3 + q_4 - q_1)) - f_T(q_1)f_{T_L}(q_3 + q_4 - q_1)f_T(-q_3)f_{T_L}(-q_4)\}. \tag{72}$$

Now integrate over $q_4$, setting $q_4 = -q_3$:

$$\frac{2\pi}{\hbar}U^2\frac{v_F^2}{|u-v_F|}\int_{\mathbb{R}^2}dq_{13}q_1\{f_T(q_3)f_{T_L}(-q_3)f_T(-q_1)f_{T_L}(q_1) - f_T(q_1)f_{T_L}(-q_1)f_T(-q_3)f_{T_L}(q_3)\}. \quad (73)$$

Assume for simplicity that $T = 0$. Then we have

$$\frac{2\pi}{\hbar}U^2\frac{v_F^2}{|u-v_F|}\left\{\int_{-\infty}^0 dq_1\int_0^\infty dq_3\frac{q_1}{1+\exp(\beta_L v_F q_3)}\frac{1}{1+\exp(-\beta_L v_F q_1)} \right.$$
$$\left. - \int_0^\infty dq_1\int_{-\infty}^0 dq_3\frac{q_1}{1+\exp(-\beta_L v_F q_3)}\frac{1}{1+\exp(\beta_L v_F q_1)}\right\}. \quad (74)$$

We can do these integrals straightforwardly, giving

$$-\frac{2\pi}{\hbar}U^2\frac{v_F^2}{|u-v_F|}\frac{\pi^2\log 2}{6\beta_L^3 v_F^3}. \quad (75)$$

Expanding the velocity prefactor, we have

$$\frac{v_F^2}{|u-v|} = \frac{v_F}{\sqrt{1+U/\pi v_F}-1} \approx 2\frac{\pi v_F^2}{U}. \quad (76)$$

Thus our answer is actually first order in $U$, as found by Yashenkin et al as well. The final answer is then (to first order in $U$)

$$\partial_t J_E(t=0) \sim U\frac{2\pi^4\log 2}{3\hbar v_F}k_B^3\left(T_R^3 - T_L^3\right), \quad (77)$$

giving a finite Fermi's Golden Rule rate $\Gamma$. For ease of reference, $2(\pi^4\log 2)/3 \approx 45$.

## IV. HIGHER DIMENSIONS AND LACK OF DIVERGENCE

In this appendix, we detail a two-dimensional version of the 1D Hubbard model quench presented in the main text, showing numerical evidence that it does not suffer from the same divergence.

Consider the two-dimensional Hubbard model,

$$H = \sum_{\langle ij\rangle,\sigma} t_{ij}c_{i,\sigma}^\dagger c_{j,\sigma} + h.c. + U\sum_i n_{i,\uparrow}n_{i,\downarrow}.$$

When $U = 0$ this model can be diagonalized by Fourier transform, giving energies

$$E = -2t_x\cos k_x - 2t_y\cos k_y, \quad (78)$$

where we have set $t_{ij} = t_x$ or $t_y$ for horizontal or vertical bonds. In what follows let us further simplify to the case $t_x = t_y = t$. We again initialize the problem with an initial $J_E^\downarrow$ as in the one-dimensional case, and at $t=0$ quench on the interactions $U$. We compute $\partial_t J_E^\uparrow$ numerically using lowest-order perturbation theory as in the main text. We note that we do not expect the same divergence in the 2D case as the 1D one due to fundamental differences in the band structure scattering processes. Our results are summarized in Fig. 1; we do not observe the same divergence in 2D as in the 1D case at the largest accessible system sizes, and the Fermi's Golden Rule limit of a constant rate appears to be valid.

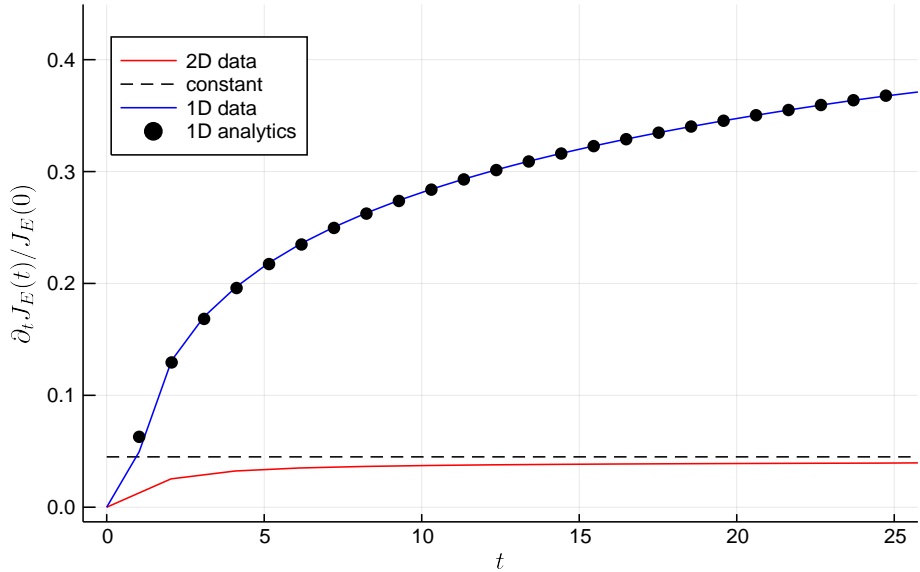

FIG. 1. Comparison of the 2D Hubbard model quench with $L_x = L_y = 100$, $t_x = t_y = t = 1.0$ (red line), to the 1D case with $L = 101$ (blue line). All data was taken at temperature $\beta_\uparrow = \beta_\downarrow = 1.0$, $U = 1.0$ and chemical potential $\mu = 0.0$ (the particle-hole symmetric point). We see that the 2D quench does not appear to be divergent, in accordance with expectations.

## V. GENERIC INTERACTIONS

In the main text, we considered Hubbard contact interactions, which are $k$-independent. In this appendix we show that the divergence associated to the response energy current occurs for generic interactions.

Let us first consider Coulomb interactions. We expect [6]

$$U_{\text{Coulomb}}(x) = U \frac{1}{\sqrt{x^2 + d^2}} \tag{79}$$

with $d$ the distance between the layers (or a cutoff for the spin-spin interactions). The Fourier transform of this is a modified Bessel function of the second kind (up to a constant prefactor),

$$U_{\text{Coulomb}}(q) = U K_0(|q|\, d). \tag{80}$$

This function is logarithmically singular near $(d\,|k|) \to 0$ and falls off exponentially as $e^{-d|k|}/\sqrt{|k|}$ at large $d\,|k|$. Following the analysis in Appendix II, we expect a log growth of the energy drag with prefactor

$$I(T) = \int_0^\pi dk K_0(kd)^2 \frac{\sin^2(k)\csc(k/2)}{1 + \cosh(2t\beta \sin(k/2))}, \tag{81}$$

which may be integrated numerically. Finally, let us consider short-ranged interactions with decay length $\xi$, i.e.

$$U_{\exp}(x) = U \frac{e^{-|x|/\xi}}{2\xi}, \tag{82}$$

which is normalized such that $\xi \to 0$ recovers the contact interaction. This has Fourier transform

$$U_{\exp}(q) = U \frac{1}{1 + \xi^2 q^2}, \tag{83}$$

entailing a prefactor of the log growth of

$$I(T) = \int_0^\pi dk \frac{1}{(1 + \xi^2 k^2)^2} \frac{\sin^2(k)\csc(k/2)}{1 + \cosh(2t\beta \sin(k/2))}. \tag{84}$$

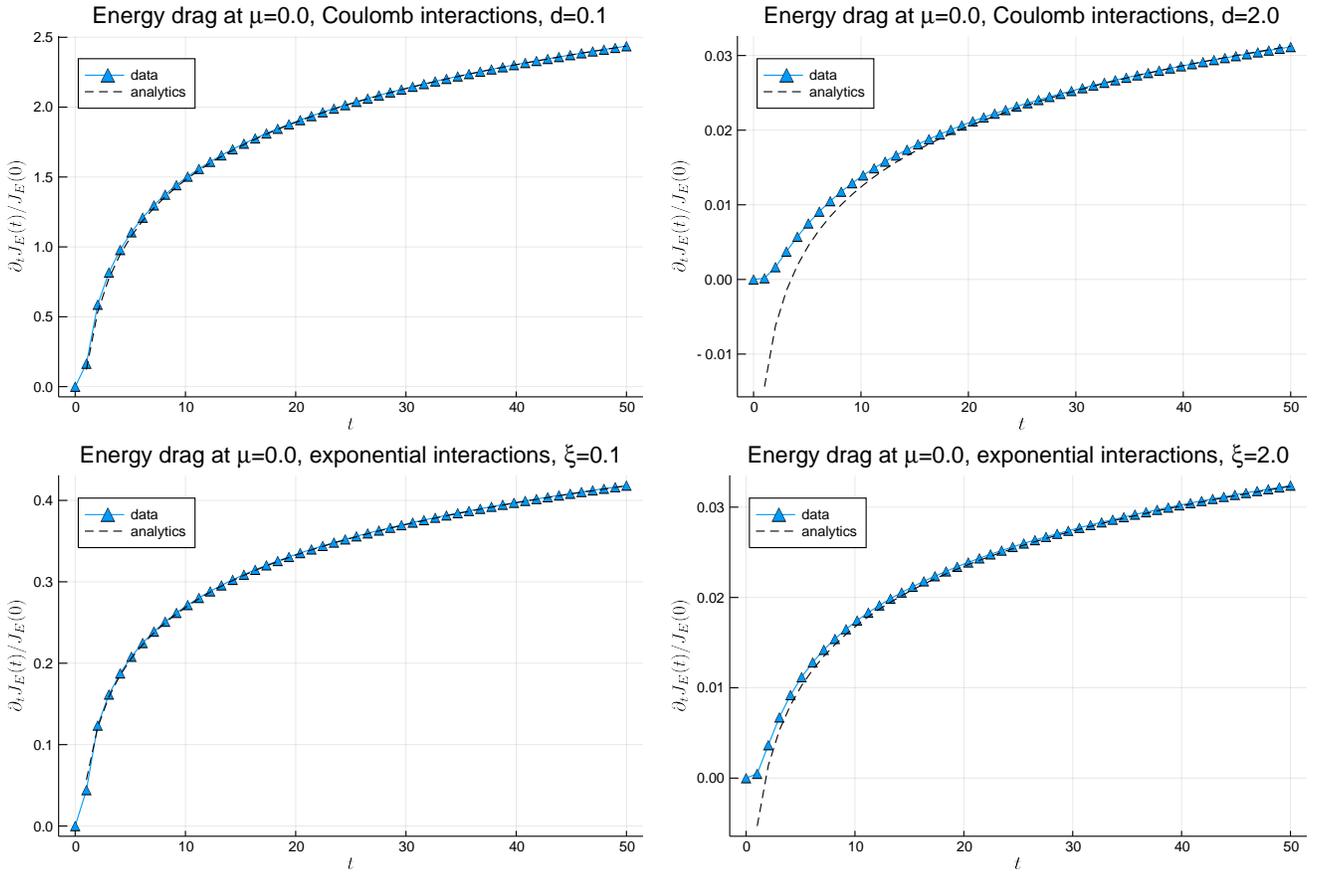

FIG. 2. Logarithmic growth of the response energy current for Coulomb interactions with $d = 0.1, 2.0$ and for short-ranged interactions with decay length $\xi = 0.1, 2.0$. All data was taken with $\mu = 0.0$, $\beta = 1.0$ with hopping parameter $t = 1.0$, on system size $L = 301$ (in the thermodynamic limit).

We have checked these predictions numerically (see Fig. 2) and find excellent agreement.

---


\newpage

\end{document}